\documentstyle{article}
\newcommand{\be}{\begin{equation}}
\newcommand{\ee}{\end{equation}}
\newcommand{\bt}{\beta}
\newcommand{\fr}{\frac}
\begin{document}
\begin{center}
{\large\bf A LINEARLY EXPANDING UNIVERSE WITH VARIABLE $\rm G$ AND
$\Lambda$ }\\
\vspace{0.8in}
{\large\bf Arbab I. Arbab\footnote{arbab64@hotmail.com}}\\
\vspace{0.3in}
{\small Department of Physics, Faculty of Science, University of
Khartoum,
Khartoum 11115, SUDAN}
\end{center}
\vspace{1in}
%\begin{abstract}
\centerline{ABSTRACT}

We have studied a cosmological model with a cosmological term of the
form
$\Lambda=3\alpha\fr{\dot R^2}{R^2}+\bt\fr{\ddot
R}{R}+\fr{3\gamma}{R^2}\ , \ \alpha, \
\bt\ , \gamma$ are constants. The scale factor (R) is found to vary
linearly
with time for both radiation and matter dominated epochs.
The cosmological constant is found to decrease as $t^{-2}$ and the rate
of
particle creation is smaller than the Steady State value.
The model gives $\Omega^\Lambda=\fr{1}{3}$ and $\Omega^m=\fr{2}{3}$ in
the present era,
$\Omega^\Lambda=\Omega^r=\fr{1}{2}$ in the radiation
era.
The present age of the universe $(\rm t_p$) is found to be $\rm
t_p=H_p^{-1}$
, where $\rm H_p$ is the Hubble constant.
The model is free from the main problems of the Standard Model. Since
the scale
factor $\rm R\propto t$ during the entire evolution of the universe the
ratio
of the cosmological constant at the Planck and present time is
$\rm\fr{\Lambda_{Pl}}{\Lambda_p}=10^{120}$.
This decay law justifies why, today, the cosmological constant is
exceedingly small.
\\
\vspace{0.5cm}
\large
\vspace{0.3in}
\\
The present interest in the flat cosmological constant models is
motivated
by the fact that a non-zero $\Lambda$ term helps to reconcile inflation
with
observation. This term could be responsible for the missing mass in the
universe.
It also helps to obtain, for a flat universe, a theoretical age  in the

observed range, even for a high value of the Hubble constant.
It has been announced by NASA group that they have found the age of the

universe to be about 12 billion years old [1].
In an attempt to reconcile this value with FRW universe, we introduce
a cosmological term ($\Lambda$) of the form
$\rm\Lambda=3\alpha\frac{\dot R^2}{R^2}+\bt\fr{\ddot
R}{R}+\fr{3\gamma}{R^2}$.
We have shown that the scale factor ($\rm R$) varies as $\rm R\propto
t$
in both radiation and matter epochs. The model has then no horizon, age
or
flatness problem. Our analysis shows that $\rm H_p=81 \ km s^{-1}
Mpc^{-1}$, a
result that is allowed by current observations. It turns out that the
$\bt$
does not affect our cosmology if the universe is singular.
%\vspace{1cm}
%\\
\newpage
{\bf 2. THE MODEL}\\

In a Robertson Walker metric, the Einstein's field equations
with  variable cosmological and gravitational `constants' and  a
perfect
fluid yield [2]
\be
\rm 3\fr{\dot R^2}{R^2}+\fr{3k}{R^2}=8\pi G\rho+\Lambda\ ,
\ee
\be
\rm 2\fr{\ddot R}{R}+\fr{\dot R^2}{R^2}+\fr{k}{R^2}=-8\pi Gp+\Lambda\ ,
\ee
where $\rho$ is the fluid energy density and $p$ its pressure.
The equation of the state is usually given by
\be
\rm p=\omega\rho\ ,
\ee
where $\omega$ is a constant. Elimination of $\rm\ddot{R}$  gives
\be
\rm
3(p+\rho)\dot{R}=-(\frac{\dot{G}}{G}\rho+\dot{\rho}+\frac{\dot{\Lambda}}{8\pi
G})R.
\ee
\vspace{0.2cm}
%\newpage
{\bf 3.a  PARTICLE CREATION}\\
\vspace{.2cm}
{\it 3.a.1 Matter-Dominated Universe}
\\
For a pressure-less ($\rm p=0$) universe, eq.(4) reads
\be
\rm\fr{d(\rho R^3)}{dt}=-\fr{R^3}{8\pi G}\fr{d\Lambda}{dt}\ .
\ee
In this paper we discuss flat cosmological models with a $\Lambda$ term

varying as [3]
\be
\rm\Lambda=3\alpha\frac{\dot R^2}{R^2}+\bt\fr{\ddot
R}{R}+\fr{3\gamma}{R^2}\ ,
\ee
where $\alpha$, $\bt$ and $\gamma$ are arbitrary constants.
This model extends the Carvalho {\it et al.} model to include the
possibility of variable gravitational constant [4].
Eqs.(2) and (6) yield
\be
\rm (2-\bt)\ddot RR+(1-3\alpha)\dot R^2-\fr{3\gamma}{R^2}=0  \ ,
\ee
which yields
\be
\rm\dot R^2=\fr{3\gamma}{(1-3\alpha)}+AR^{-2(1-3\alpha)/(2-\bt)}\ ,
\ee
where $\rm A= constant $.
If A=0 (singular solution), one obtains,
\be
\rm R=\sqrt{\fr{3\gamma}{(1-3\alpha)}}\ t\ ,\ \alpha<\fr{1}{3}\ .
\ee
Eqs.(1), (6) and (9)  yield
\be
\rm H_p=\fr{1}{t_p}\ ,\ \Lambda(t)=\fr{1}{t_p^2}\ ,\ \rho=\fr{1}{4\pi
G}\fr{1}{t_p^2}\  .
\ee
where $\rm H=\fr{\dot R}{R}$ is the Hubble constant. Hereafter the
subscript
`p' denotes the present value of the quantity.
The vacuum energy density ($\rho_v$) is given by
\be
\rm\rho_v(t)=\fr{\Lambda}{8\pi G}=\fr{1}{8\pi G}\fr{1}{t^2}\ .
\ee
The deceleration parameter ($\rm q$) is defined as
\be
\rm q\equiv -\fr{\ddot RR}{\dot R^2}=0
\ee
The density parameter of the universe ($\rm\Omega^m$) is given by
\be
\rm\Omega^m=\fr{\rho}{\rho_c}=\fr{2}{3}\ ,
\ee
where $\rm\rho_c=\fr{3H^2}{8\pi G}$.
The density parameter due to vacuum contribution is defined as
$\rm\Omega^\Lambda=\fr{\Lambda}{3H^2}$.
Using eq.(10) this yields
\be
\rm\Omega^\Lambda=\fr{1}{3}\ .
\ee
It is interesting that the above cosmological parameters are
independent of
the value of $\alpha, \bt$ or $\gamma$.
We shall define $\rm\Omega_{\rm total}$ as
\be
\rm\Omega_{\rm total}=\Omega^m+\Omega^\Lambda\ .
\ee
Hence eqs.(13), (14) and (15) give $\rm\Omega_{\rm total}=1$.
The  inflationary paradigm requires this solution.
A very recent age of the universe is shown to be 12 billion years. This
would imply that the Hubble constant $\rm H_p = 81\ kms^{-1}Mpc^{-1}$.
This value is consistent with the recent Hubble Space Telescope
determination
of $\rm h=0.80\pm0.17$ [5].
We now turn to calculate the rate of particle creation (annihilation)
$\rm n$, which is defined as [6]
\be
\rm n=\fr{1}{R_p^3}\fr{d(\rho R^3)}{dt}|_p\ .
\ee
Using eqs.(5), (9) and (10),  eq.(16) yields
\be
\rm n_p=\rho_p H_p \ .
\ee
We remark that this rate is less than that of the Steady State
model ($\rm=3\rho_p H_p$).
\\

{\bf A model with variable $\rm G$}:
\\

We now consider a model in which both $G$ and $\Lambda$ vary with time
in such a way  the usual energy conservation law holds.
Equation (4) can be split to give [2]
\be
\rm\dot\rho+3 H\rho=0\ ,
\ee
and
\be
\rm\dot\Lambda+8\pi\dot G\rho=0\ .
\ee
Using eqs.(9) and (10), eqs.(18) and (19) yield
\be
\rm\rho(t)=A't^{-3}\ ,
\ee
and
\be
\rm G(t)=\fr{1}{4\pi G A'}\ t  \ , \ A'=const.
\ee
This solution is obtained by Berman and Som [7] and can be obtained
from
Berman's model [8] with $[m=1, \alpha=0$].
It has been shown that the development of the large-scale anisotropy
is given by the ratio of the shear $\sigma$ to the volume expansion
($\theta=3\fr{\dot R}{R}$) which evolves as [9]
\be
\rm\fr{\sigma}{\theta}\propto t^{-2}\ .
\ee
The present observed isotropy of the Universe requires this anisotropy
to be decreasing as the universe expands. Thus an increasing $G$
guarantees
an isotropized universe.
\\
\vspace{.2cm}

{3.a.2 \it Radiation-Dominated Universe}
\\

This is defined by the equation of the state $\rm p=\frac{1}{3}\rho\
(\omega=\fr{1}{3})$.
Eqs.(1), (2) and (6) yield
\be
\rm (1-2\alpha)\frac{\dot R^2}{R^2}+(1-\fr{2}{3}\bt)\frac{\ddot
R}{R}-\fr{2\gamma}{R^2}=0\ ,
\ee
which gives
\be
\rm\dot
R^2=\fr{2\gamma}{(1-2\alpha)}+BR^{-2(1-2\alpha)/(1-\fr{2}{3}\bt)}\ ,
\ee
where B=const. For B =0 (singular solution), one obtains
\be
\rm R=\sqrt{\fr{2\gamma}{(1-2\alpha)}}\ t \ ,\ \alpha<\fr{1}{2}\ .
\ee
Eqs.(1) and (6) give
\be
\rm\Lambda=\fr{3}{2}\fr{1}{t^2}\ , \
\rm\rho=\fr{3}{16\pi G}\fr{1}{t^2}\ .
\ee
It has been pointed out that we need not have an inflationary phase
(exponential
law for $\rm R$) because there is no horizon problem with the above
solution [8].
We see from eqs.(10) and (26) that the ration of the cosmological
constant at
the Planck's time ($\rm t_{Pl}=10^{-43}\ sec$) and the present present
time ($\rm t_p=10^{17}\ sec)$ is
$\rm\fr{\Lambda_{Pl}}{\Lambda_p}=(\fr{10^{17}}{10^{-43}})^2=10^{120}$.
Thus $\Lambda$
,today, is exceedingly small because the universe is too old!.
Eqs.(15) and (26) yield
\be
\rm\Omega^r=\Omega^\Lambda=\fr{1}{2}\ .
\ee
Again we see that the above cosmological parameters are independent of
the
value of $\alpha,\ \bt$ or $\gamma$. Equation (27) shows that the
radiation and
vacuum contribute equally to the total energy density.
%\\
%\vspace{0.2cm}
\newpage
{\bf A model with variable $\rm G$}:
\\
Equation (4) now reads
\be
\rm\dot\rho+4H\rho=0\ ,
\ee
and
\be
\rm\dot\Lambda+8\pi \dot G\rho=0\ .
\ee
Employing eqs.(1) and (6), eqs.(28) and (29) yield
\be
\rm\rho=B't^{-4}\ \ , \ \ G=\fr{3}{16\pi B'}t^2\ ,\ \ B'=const.
\ee
Recently, Abdel Rahman [2] and Berman [7,8,10] (with $[m=1,
\alpha=\fr{1}{3}$]) have
found that  $\rm R\propto t\ , G\propto t^{2}\ ,  \rho\propto t^{-4}$
in the
radiation era.
\\
\vspace{0.2cm}

{\bf CONCLUSION}
\\

\vspace{0.1cm}
We have analyzed in this paper the effect of the assumed decay law for
the
cosmological constant. The only possible solution is the $R\propto t$
for
the radiation and matter dominated epochs.
This solution is obtained by Berman (for Brans-Dicke models
with a time dependent cosmological term) for the matter and the
radiation
epochs and by Abdel Rahman for the radiation epoch.
There is no horizon or age problem associated with this solution.
Our model predicts that $\rm H_p=81\ km s^{-1} Mpc^{-1}$ if the age of
the universe
is 12 billion years.
Thus by introducing the cosmological constant it is possible for a flat
model
to be in accordance with the universe age even with high values of $\rm
H_p$.
We stress that the decay law $\Lambda \propto t^{-2}$
indeed plays an important role in cosmology as remarked by Berman.
A natural extension of this work would be to investigate different
scenarios
by considering  non-singular solutions, i.e., solutions with $A$ and
$B$ negative.
\\
\vspace{0.2cm}

{\bf ACKNOWLEDGMENT}
\vspace{0.2in}\\
I would like to thank the University of Khartoum for financial support.
\\
\vspace{0.2in}

{\bf REFERENCES}
%\end{center}
\vspace{0.1in}
\\
1- NASA GROUP of astronomers, 25 May 1999 \\
2- Abdel-Rahman, A. -M. M., 1990. {\it Gen. Rel. Gravit.}{\bf 22},
655\\
3- Al-Rawaf, A. S., 1998. {\it Mod. Phys. Lett.}{\bf A 13}, 429\\
4- Carvalho, J. C., Lima, J. A. S., and Waga, I., 1992. {\it Phys.
Rev.}{\bf D46}, 2404\\
5- Freedman, W. L., 1994. {\it Nature}{\bf 371}, 757\\
6- Matyjasek, J., 1995. {\it Phys. Rev.}{\bf D51}, 4154\\
7- Berman, M. S., and Som, M. M., 1990. {\it Int. J. Theor. Phys.} {\bf
29}, 1411\\
8- Berman, M. S., 1991. {\it Gen. Rel. Gravit.}{\bf 23}, 465\\
9- Barrow, J. D., 1978. {\it Mon. Not. astr. Soc.} {\bf 184}, 677\\
10- Berman, M. S., 1990. {\it Int. J. Theor. Phys.} {\bf 29}, 1419\\
\end{document}